\documentclass{aip-cp}

\usepackage[numbers]{natbib}
\usepackage{rotating}
\usepackage{graphicx}

\begin{document}

% Journal definitions
\let\jnl\rm
\def\refa@jnl#1{{\jnl#1}}
\def\aj{\refa@jnl{AJ}}                   % Astronomical Journal
\def\actaa{\refa@jnl{Acta Astron.}}      % Acta Astronomica
\def\araa{\refa@jnl{ARA\&A}}             % Annual Review of Astron and Astrophys
\def\apj{\refa@jnl{ApJ}}                 % Astrophysical Journal
\def\apjl{\refa@jnl{ApJ}}                % Astrophysical Journal, Letters
\def\apjs{\refa@jnl{ApJS}}               % Astrophysical Journal, Supplement
\def\ao{\refa@jnl{Appl.~Opt.}}           % Applied Optics
\def\apss{\refa@jnl{Ap\&SS}}             % Astrophysics and Space Science
\def\aap{\refa@jnl{A\&A}}                % Astronomy and Astrophysics
\def\aapr{\refa@jnl{A\&A~Rev.}}          % Astronomy and Astrophysics Reviews
\def\aaps{\refa@jnl{A\&AS}}              % Astronomy and Astrophysics, Supplement
\def\azh{\refa@jnl{AZh}}                 % Astronomicheskii Zhurnal
\def\baas{\refa@jnl{BAAS}}               % Bulletin of the AAS
\def\bac{\refa@jnl{Bull. astr. Inst. Czechosl.}}
                % Bulletin of the Astronomical Institutes of Czechoslovakia 
\def\caa{\refa@jnl{Chinese Astron. Astrophys.}}
                % Chinese Astronomy and Astrophysics
\def\cjaa{\refa@jnl{Chinese J. Astron. Astrophys.}}
                % Chinese Journal of Astronomy and Astrophysics
\def\icarus{\refa@jnl{Icarus}}           % Icarus
\def\jcap{\refa@jnl{J. Cosmology Astropart. Phys.}}
                % Journal of Cosmology and Astroparticle Physics
\def\jrasc{\refa@jnl{JRASC}}             % Journal of the RAS of Canada
\def\memras{\refa@jnl{MmRAS}}            % Memoirs of the RAS
\def\mnras{\refa@jnl{MNRAS}}             % Monthly Notices of the RAS
\def\na{\refa@jnl{New A}}                % New Astronomy
\def\nar{\refa@jnl{New A Rev.}}          % New Astronomy Review
\def\pra{\refa@jnl{Phys.~Rev.~A}}        % Physical Review A: General Physics
\def\prb{\refa@jnl{Phys.~Rev.~B}}        % Physical Review B: Solid State
\def\prc{\refa@jnl{Phys.~Rev.~C}}        % Physical Review C
\def\prd{\refa@jnl{Phys.~Rev.~D}}        % Physical Review D
\def\pre{\refa@jnl{Phys.~Rev.~E}}        % Physical Review E
\def\prl{\refa@jnl{Phys.~Rev.~Lett.}}    % Physical Review Letters
\def\pasa{\refa@jnl{PASA}}               % Publications of the Astron. Soc. of Australia
\def\pasp{\refa@jnl{PASP}}               % Publications of the ASP
\def\pasj{\refa@jnl{PASJ}}               % Publications of the ASJ
\def\rmxaa{\refa@jnl{Rev. Mexicana Astron. Astrofis.}}%
                % Revista Mexicana de Astronomia y Astrofisica
\def\qjras{\refa@jnl{QJRAS}}             % Quarterly Journal of the RAS
\def\skytel{\refa@jnl{S\&T}}             % Sky and Telescope
\def\solphys{\refa@jnl{Sol.~Phys.}}      % Solar Physics
\def\sovast{\refa@jnl{Soviet~Ast.}}      % Soviet Astronomy
\def\ssr{\refa@jnl{Space~Sci.~Rev.}}     % Space Science Reviews
\def\zap{\refa@jnl{ZAp}}                 % Zeitschrift fuer Astrophysik
\def\nat{\refa@jnl{Nature}}              % Nature
\def\iaucirc{\refa@jnl{IAU~Circ.}}       % IAU Cirulars
\def\aplett{\refa@jnl{Astrophys.~Lett.}} % Astrophysics Letters
\def\apspr{\refa@jnl{Astrophys.~Space~Phys.~Res.}}
                % Astrophysics Space Physics Research
\def\bain{\refa@jnl{Bull.~Astron.~Inst.~Netherlands}} 
                % Bulletin Astronomical Institute of the Netherlands
\def\fcp{\refa@jnl{Fund.~Cosmic~Phys.}}  % Fundamental Cosmic Physics
\def\gca{\refa@jnl{Geochim.~Cosmochim.~Acta}}   % Geochimica Cosmochimica Acta
\def\grl{\refa@jnl{Geophys.~Res.~Lett.}} % Geophysics Research Letters
\def\jcp{\refa@jnl{J.~Chem.~Phys.}}      % Journal of Chemical Physics
\def\jgr{\refa@jnl{J.~Geophys.~Res.}}    % Journal of Geophysics Research
\def\jqsrt{\refa@jnl{J.~Quant.~Spec.~Radiat.~Transf.}}
                % Journal of Quantitiative Spectroscopy and Radiative Transfer
\def\memsai{\refa@jnl{Mem.~Soc.~Astron.~Italiana}}
                % Mem. Societa Astronomica Italiana
\def\nphysa{\refa@jnl{Nucl.~Phys.~A}}   % Nuclear Physics A
\def\physrep{\refa@jnl{Phys.~Rep.}}   % Physics Reports
\def\physscr{\refa@jnl{Phys.~Scr}}   % Physica Scripta
\def\planss{\refa@jnl{Planet.~Space~Sci.}}   % Planetary Space Science
\def\procspie{\refa@jnl{Proc.~SPIE}}   % Proceedings of the SPIE

\let\astap=\aap
\let\apjlett=\apjl
\let\apjsupp=\apjs
\let\applopt=\ao

% ==== Actual document start ==== %

\title{The Use Case of a New ISRF on Diffuse Gamma-ray Emission Models}

\author[aff1]{Felix Niederwanger}
\eaddress{felix.niederwanger@uibk.ac.at}
\author[aff1]{Olaf Reimer}
\author[aff1]{Ralf Kissmann}
\author[aff4]{Richard Tuffs}

\affil[aff1]{Institut f\"ur Astro- und Teilchenphysik, Leopold-Franzens-Universit\"at Innsbruck, A-6020 Innsbruck, Austria}
\affil[aff4]{Max Planck Institut für Kernphysik}

\maketitle

\begin{abstract}
An important contribution to the gamma-ray emission of our Galaxy is the Galactic Diffuse emission.
We present specific developments within the PICARD code for modeling of Galactic cosmic ray propagation that are relevant for the computation of the gamma-ray emission to accommodate a new Interstellar Radiation Field (ISRF) model.
We study the differences of the individual ISRF models on Galactic Diffuse gamma-ray emission predictions in the GeV to the TeV energy regime.

Emphasis was laid on obtaining robust predictions for observable signatures in the very high energy gamma-ray regime with a special attention to the energy regime for H.E.S.S.
Preliminary longitude-latitude profile and residuum maps of inverse Compton emissions are shown for axisymmetric and four-arm spiral Galaxy models.

\end{abstract}

\section{INTRODUCTION}

The Galactic Diffuse emission is a prominent signature of gamma-emission in the regime above 100 MeV \cite{2012ApJ...750....3A}. It is produced via cosmic-ray interactions on the interstellar gas distribution and interstellar radiation fields (ISRF).
One of the recent extension to the PICARD code \cite{2014APh....55...37K} is the implementation of a new ISRF, based on observational data compiled by Tuffs and Popescu.
The data of the ISRF model relates to two components: Starlight originating from stellar populations, and scattered light by dust. Both components use a different frequency binning and resolution and are dominant in different wavelength regimes, thus allowing us to generate a full frequency range spectrum ISRF for the whole Milky-way by adding the contributions of the cosmic microwave background.

The use case of the new ISRF is to create all-sky gamma-ray flux predictions using the PICARD code. We present the implementation details, the
structure of the new ISRF including its spectral shape, and predictions in the inverse Compton (IC) channel for the gamma-ray skymaps, where the most dominant features are to be expected. Additional contributions are expected also in the intensities of gamma rays produced by Bremsstrahlung because of differences in the electron distribution within the Galaxy, due to changes in the energy-loss rates of electrons.

Preliminary longitude and latitude profiles have been produced with the frequency binning of the ISRF model and indicate differences for the inverse Compton gamma-ray flux at different energy regimes.

\section{Structure of the ISRF}

\begin{figure}[h]
  \centerline{\includegraphics[width=\textwidth]{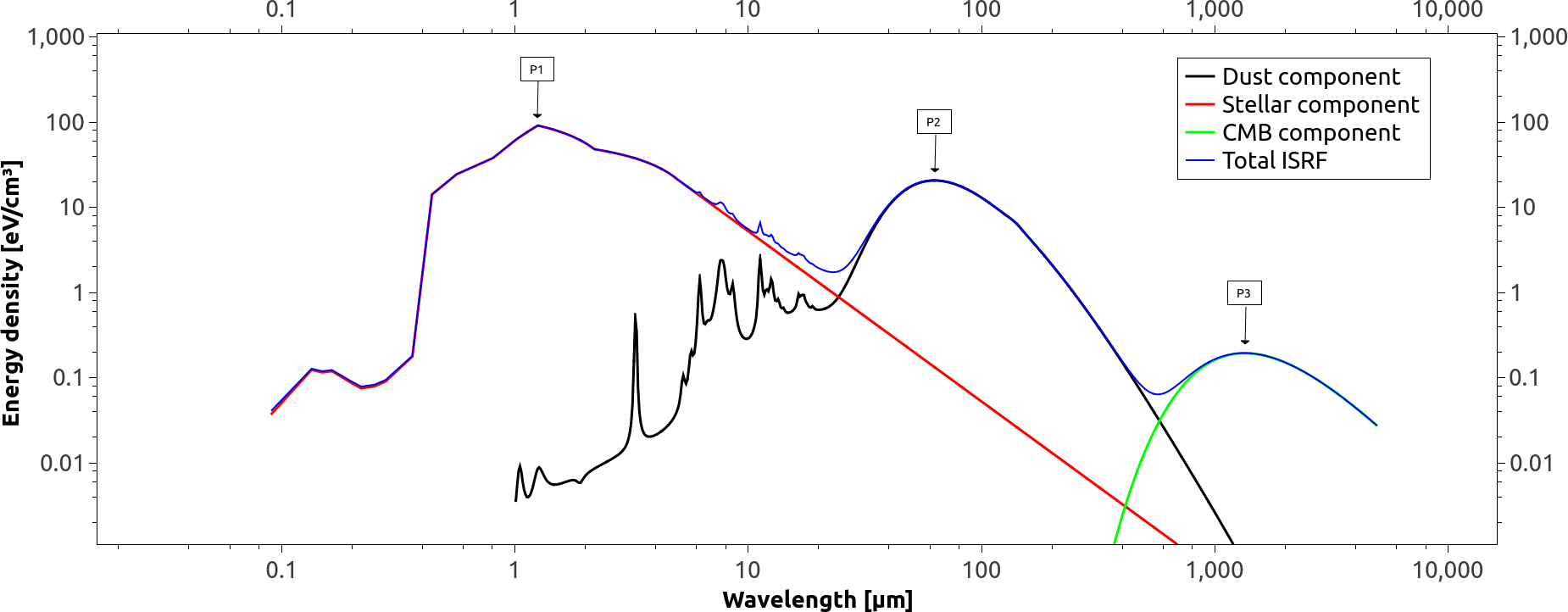}}
  \caption{Energy spectrum of the different components in the ISRF at the Galactic center}
  \label{fig:ISRFSpectrum}
\end{figure}

The ISRF we use in PICARD is given as an axisymmetric distribution at different frequencies. The frequency binning is chosen to accurately resolve the most relevant features of the spectrum.
The ISRF consists of two components: Starlight from the stellar population (approximately $0.1\,-10\,\mu m$) and the light scattered by the dust in the Galaxy in the range around $1\,\mu m - 1\,mm$. An additional component added within PICARD but not present in the compiled ISRF model by Popescu and Tuffs is the cosmic microwave background (CMB), dominating the other components above roughly $400\,\mu m$. An intensity spectrum of the different components is shown in figure \ref{fig:ISRFSpectrum}.

We have data for the starlight component up to approximately $10\,\mu m$, but starlight is still the dominant part of the ISRF up to $20\,\mu m$, where dust becomes the most prominent component. To extend the starlight component to higher wavelengths we use the Rayleight-Taylor approximation.

\begin{equation}
	B_\nu(\nu,T) = \frac{2\nu^2}{c^2}kT
	\label{eq:RayleighJeans}
\end{equation}

This approach is valid, since we are far away from the UV regime.

We use extrapolation for the dust component in order to extend it to the starlight and CMB regimes.
The dust regime is defined for wavelengths between $1\,\mu m$ and $5\,mm$. Beyond this wavelength range the contribution of dust to the total ISRF is dominated by starlight towards lower wavelengths and by the CMB towards higher wavelengths. For wavelengths below $1 \, \mu m$ the energy density of the dust component is negligible compared to the starlight. The same applies for the CMB-dominated regime above $400-800\,\mu m$, where the dust and starlight component are also at least one order of magnitude below the CMB.

\begin{figure}[h]
  \centerline{\includegraphics[width=\textwidth]{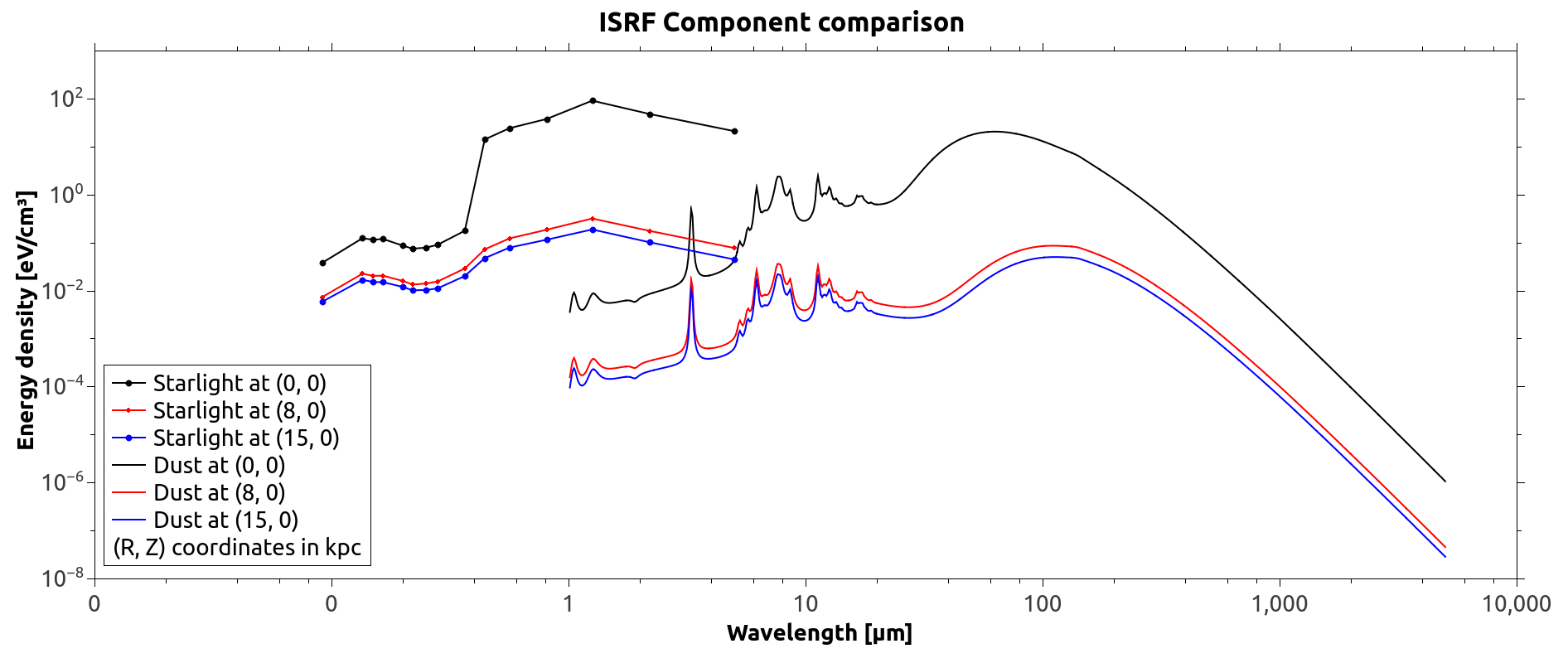}}
  \caption{Starlight and dust component of the new ISRF at different $(R, z)$ locations within the Galaxy}
  \label{fig:ISRFComponents}
\end{figure}

The starlight and dust components at different locations in the Galaxy are shown in figure \ref{fig:ISRFComponents}. The transition from UV to optical light of the stellar component at roughly $0.4\,\mu m$ has a discontinuity at the $(0,0)$ coordinate and is smooth outside of the Galactic center.

The starlight component is defined over 15 frequency bins while the dust component has 629 of them. The dust component has much more frequency bins in order to resolve the given spectral features accurately (see figure \ref{fig:ISRFComponents}).
 
In comparison to the GALPROP ISRF v3 \cite{2008ApJ...682..400P}, these spectral features are now available in a necessary accuracy and resolution for upcoming simulations of cosmic-ray transport and the compuation of gamma-ray emission using PICARD. For ease of comparison we use the same frequency binning as GALPROP v3 in order to compare the ISRF at given frequencies. We expect additional deviations when using the full new frequency range.

\subsection{PICARD - The simulation framework}

The PICARD code consists of two successive parts: The propagation of cosmic rays and the computation of gamma-ray emission \cite{2014APh....55...37K}.

The transport physics can be defined via a set of parameters. We have computed gamma-ray emissions for two different models: One with an axisymmetric source distribution and one with a four-arm spiral Galaxy model \cite{2015APh....70...39K,2014ApJS..215....1V,2010ApJ...722.1460S}.
The source distribution for the axisymmetric model is based on the radial distribution of pulsars within the Galaxy by Yusifov and K\"uc\"uk \cite{2004A&A...422..545Y}.
For a better comparison between the differences of the two ISRF models we use a 4-arm spiral Galaxy source model where all parameters except the source distribution are the same as for the axisymmetric model. This does not represent the comsic-ray fluxes at the position of Earth accurately, some propagation parameters would need to be modified.

The output of the first step (Propagation) is a converged solution of the cosmic-ray densities of the different species across the whole Galaxy. It is stored in HDF5 files, that can be used for individual analysis of the cosmic-ray data or as input for the upcoming step:
The computation of the gamma-ray emission, using the pre-calculated fluxes of cosmic rays.
The source distribution of cosmic rays within the Galaxy influences the converged solution of their densities, and this is visible in the resulting gamma-ray emissions, especially in the Galactic plane, where spiral arm tangents can be seen.
There are three channels relevant for the production of gamma-rays: Decay of neutral pions, Bremsstrahlung and inverse Compton scattering.
The decay of netral pions ($\pi^0$) and Bremsstrahlung are computed via the scattering of the ISM gas within the Galaxy. This means, that a new ISRF will not change those two production channels directly. Indirectly, Bremsstrahlung is modified by a different ISRF, because the distribution of electrons in the Galaxy change due to different energy loss processes.
The only gamma-ray production channel used in PICARD and directly influenced by the new ISRF is the inverse Compton scattering of leptons.

\section{Predictions of the gamma-ray fluxes}

Using the new ISRF we computed all-sky gamma-ray flux predictions for the different Galactic models: the axisymmetric and the four-arm model.
The chosen energies are 10 GeV, 1 TeV and 100 TeV. In the high TeV regime, especially at 100 TeV, the assumption of a smooth cosmic-ray source distribution is questionable and may be replaced in the future by a catalog of individual sources, once available. For the Galaxy we choose an axi-symmetric source distribution and a four-arm spiral Galaxy model \cite{2015APh....70...39K}.

\begin{figure}[h]
  \centerline{\includegraphics[width=\textwidth]{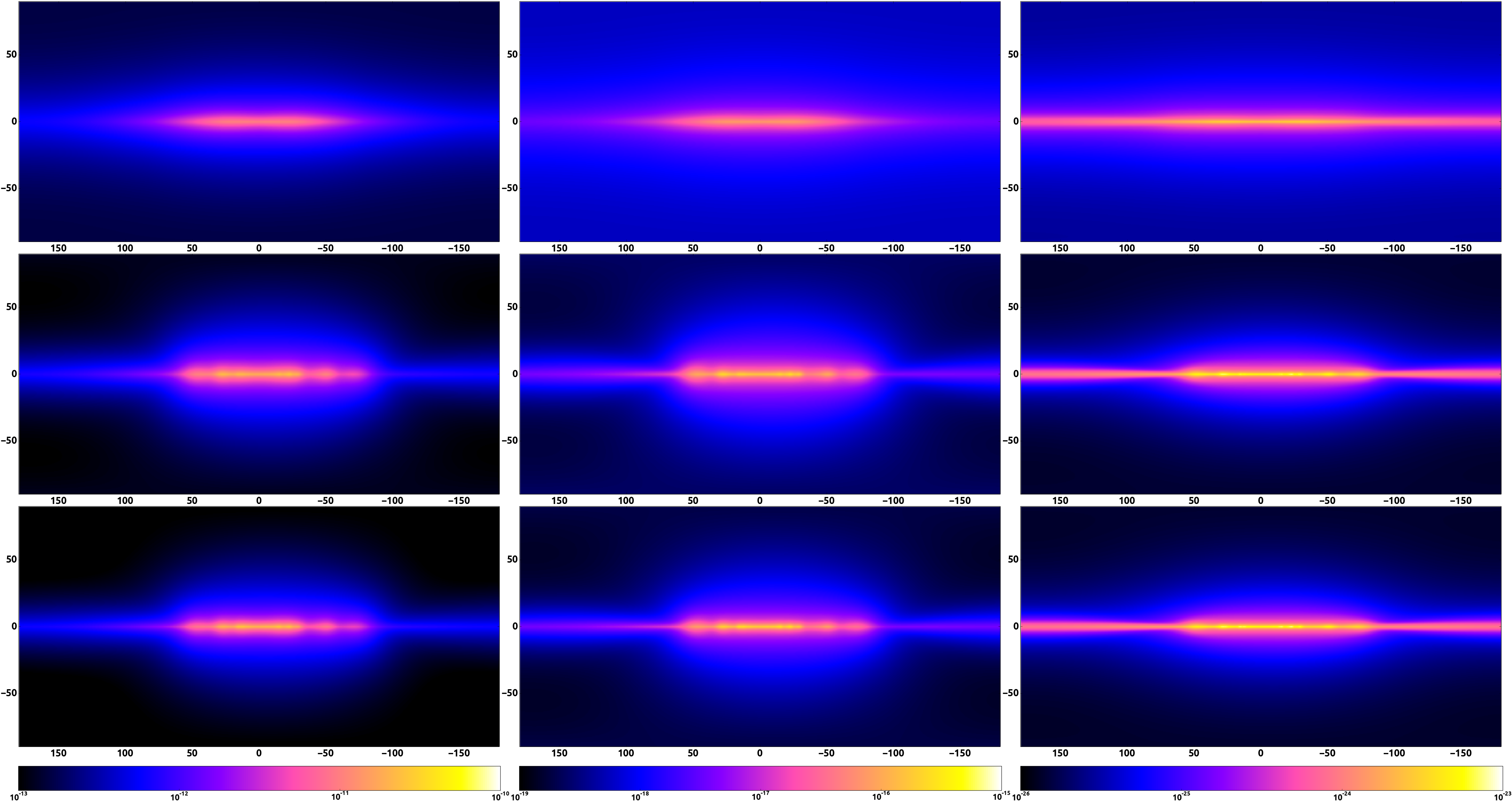}}
  \caption{Predictions of IC channel of the gamma-ray fluxes at 10 GeV (left), 1 TeV (middle) and 100 TeV (right) for an axi-symmetric model (top), a 4-arm spiral Galaxy model \cite{2015APh....70...39K} with the GALPROP v3 radiation field \cite{2008ApJ...682..400P} (middle) and a 4-arm spiral Galaxy model with the new ISRF (bottom)}
  \label{fig:GammaMaps}
\end{figure}

While the axi-symmetric distribution shows a smooth and symmetrical distribution as expected, the four-arm spiral Galaxy models show additional details. One can find more morphological structures especially in the inner Galaxy, where the spiral arm tangents can be seen. Furthermore the flux at higher latitudes is reduced, especially with the new ISRF.

\begin{figure}[h]
  \centerline{\includegraphics[width=0.8\textwidth]{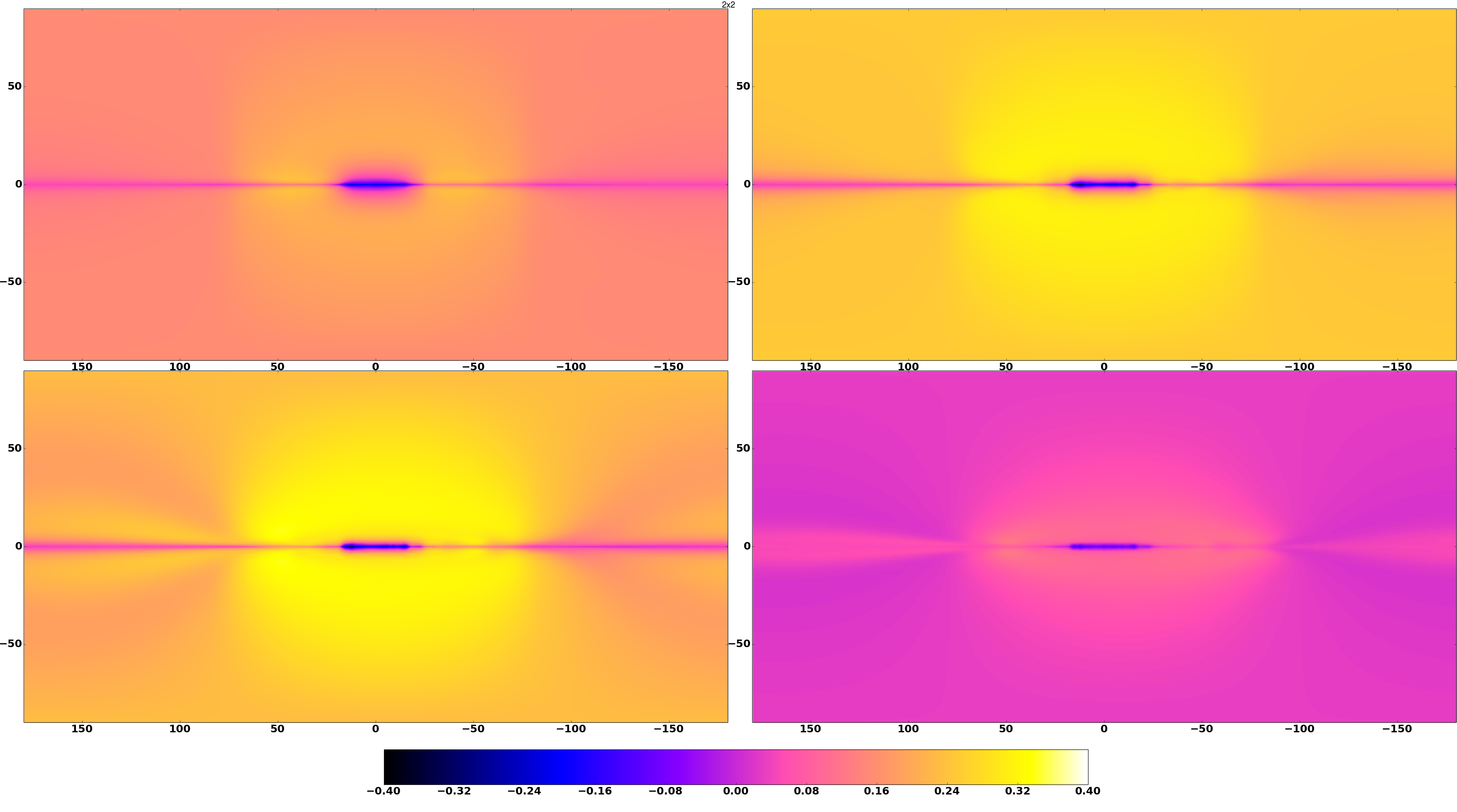}}
  \caption{Residuum maps ($(A-B)/A$) of the IC component for 100 MeV (top left), 100 GeV (top right), 1 TeV (bottom left) and 100 TeV (bottom right). The GALPROP v3 ISRF ($A$) is compared to the new ISRF ($B$), both using the same underlying 4-arm spiral Galaxy model.}
  \label{fig:ResiduumICS}
\end{figure}

Figure \ref{fig:GammaMaps} shows different gamma-ray fluxes at the following energies: 10 GeV (left), 1 TeV (middle) and 100 TeV (right) for three different models: On top we have an axi-symmetric model, in the middle there is a four-arm spiral Galaxy model \cite{2015APh....70...39K} with the GALPROP v3 ISRF \cite{2008ApJ...682..400P} and on the bottom we have a four-arm spiral Galaxy model with new PICARD ISRF. For this
work, the PICARD ISRF was downsampled to the same frequency bins as the GALPROP v3 model.
In the case of the four-arm spiral Galaxy model, the new ISRF reduces the intensity of the
inverse Compton emission outside of the Galactic center, especially for the MeV and the high GeV energies. In the Galactic center the gamma-ray flux is higher when using the
new ISRF model. Figure \ref{fig:ResiduumICS} shows all-sky relative residuum maps ($\frac{A-B}{A}$) for 100 MeV, 100 GeV, 1 TeV and 100 TeV inverse Compton emission using a 4-arm spiral Galaxy model and compares the GALPROP v3 ISRF ($A$) to the new ISRF ($B$).

\begin{figure}[h!]
  \centerline{\includegraphics[width=0.8\textwidth]{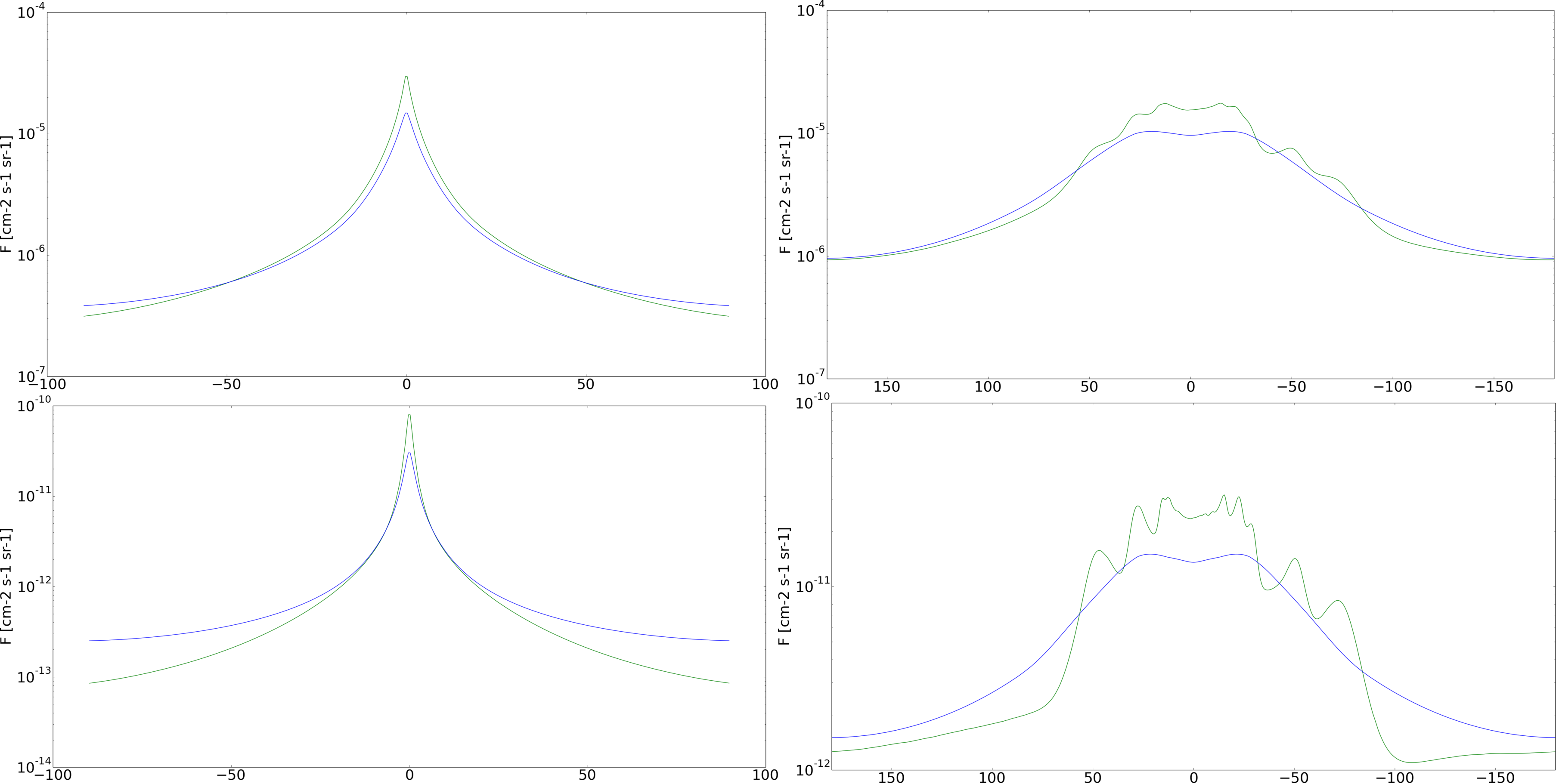}}
  \caption{Latitude profile of an axisymmetric model with the GALPROP v3 ISRF (blue) and a four-arm spiral Galaxy model with the new ISRF (green). The plots show the integrated energy density over the range of 10 GeV - 1 TeV.}
  \label{fig:LonLatProfile2}
\end{figure}

\begin{figure}[h!]
  \centerline{\includegraphics[width=0.8\textwidth]{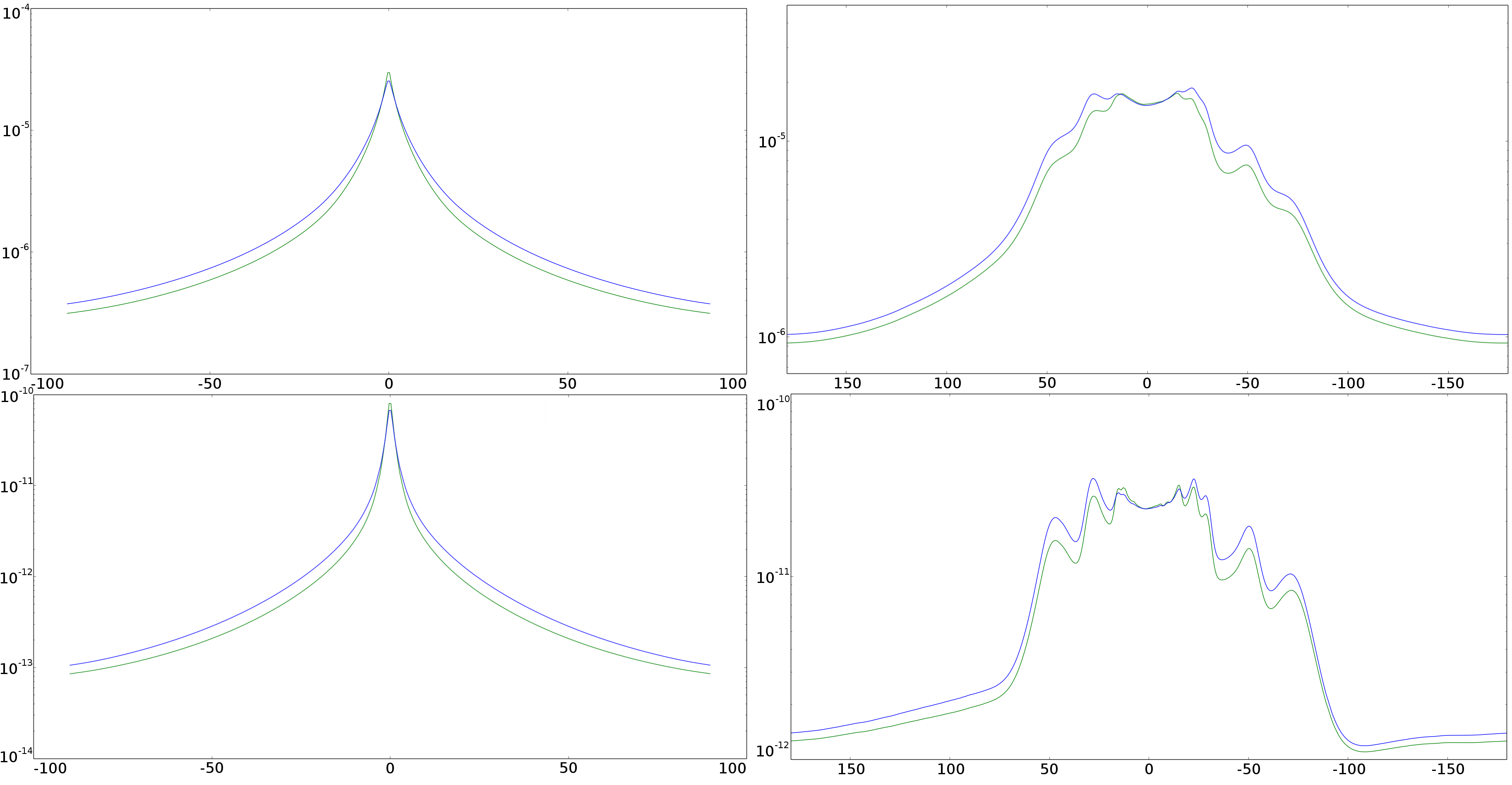}}
  \caption{Latitude (left) and longitude (right) profile plots for integrated energy densities from 100 MeV to 10 GeV (top) and 1 TeV to 100 TeV (bottom). Longitude and latitude profiles are averaged within $5$ degrees from the center ($|b| <= 5 ^\circ$ and $|l| <= 5 ^\circ$). The underlying model was the four-arm spiral Galaxy model with the GALPROP v3 ISRF (blue) and the new ISRF (green)}
  \label{fig:LonLatProfile}
\end{figure}

With respect to the axisymmetric model, the gamma-ray emission of the four-arm spiral Galaxy model is more constraint to the Galactic plane, and the spiral arm tangents become visible. They can also be identified in the latitude profiles of the axisymmetric and of the four-arm spiral Galaxy model as shown in figure \ref{fig:LonLatProfile2}.
The spiral arm tangents are visible in the four-arm models with both ISRFs.
At 1 TeV, the new ISRF decreases also the flux in the outer Galaxy with respect to the
same four-arm model but with the GALPROP v3 ISRF. This can be seen as well in the latitude profiles of the different ISRF models, shown in figure \ref{fig:LonLatProfile}.
At 100 TeV, the inverse Compton component is narrower to the galactic plane. 

Figure \ref{fig:LonLatProfile} shows preliminary central longitude and latitude profiles. The longitude profiles are computed as averages over the latidues from $-5^\circ$ to $5^\circ$, the latitude profile as averages over the longitudes from $-5^\circ$ to $5^\circ$ at different energy ranges, over witch the integrated flux has been calculated.
We can see that in general the GALPROP ISRF v3 yields the higher flux except for the Galactic bulge.
This is visible for the longitude profiles, where the new ISRF yields higher flux values only in the Galactic center, between the Norma and the Sagittarius tangents.
Longitude and latitude plots are done with the finer frequency binning mentioned before. We compare the two different ISRF on the basis of the same underlying four-arm spiral Galaxy
model, while in Figure \ref{fig:LonLatProfile2} an axisymmetric model using the GALPROP v3 ISRFs is compared to the four-arm spiral Galaxy model using the new ISRF.
In comparison to the GALPROP v3 ISRF we notice smaller gamma-ray fluxes except for the Galactic bulge, where our ISRF model increases the flux. The morphology of the inverse Compton emission component is more structured as can be seen in figure \ref{fig:ResiduumICS}, where residuum maps of the IC component are shown for the above mentioned models.
The largest differences in the four-arm spiral Galaxy model between the GALPROP v3 ISRF and our new ISRF are in the Galactic bulge, where our new ISRF yields significantly higher gamma-ray intensities.
Although we can already see important differences when applying a new ISRF, further minor deviations are to be expected due to the different frequency binning.

Summarising, the Galactic model has a greater influence on the gamma-ray emission maps than the new ISRF. In the longitude profiles in Figure \ref{fig:LonLatProfile2} we can identify the spiral arm tangents as significant morphological structure in different energy regimes.

This finding is based on a global assessment, whereas studies on local discrepancies are still to be carried out.

\section{ACKNOWLEDGMENTS}

The computational results presented have been achieved (in part) using the HPC infrastructure LEO and MACH of the University of Innsbruck. \\
The financial support from the Austrian Science Fund (FWF) project number
\textit{I1345}, in collaboration with the French Science Fund (ANR),
project ID \textit{ANR-13-IS05-0001} is acknowledged.

% References

\nocite{*}
\bibliographystyle{aipnum-cp}%

\end{document}